\documentclass[prd,showpacs,noshowkeys,floatfix,nofootinbib,fleqn,tightenlines, 
preprint]{revtex4}

\usepackage{amsmath}
\usepackage{amsfonts}
\usepackage{latexsym}
\usepackage{graphicx}
\usepackage{mathrsfs}
\usepackage{dcolumn}
\usepackage{xspace}                       

\newcommand{\version}{v4}

\newcommand{\DS}[1]{$\mathsf{#1}$\xspace}

\newcommand{\fracnew}[2]
           {\protect\frac{{#1}_{\protect\vphantom{!_a}}}{{#2}^{\protect\vphantom{a}}}}

\newcommand{\beq}{\begin{equation}}
\newcommand{\eeq}{\end{equation}}
\newcommand{\beqa}{\begin{eqnarray}}
\newcommand{\eeqa}{\end{eqnarray}}

\renewcommand{\i}{\mathrm{i}}

\newcommand{\threemat}[1]{\left( \begin{array}{ccc} #1 \end{array} \right)}

\newcommand{\SM}   {Standard Model}

\begin{document}
\noindent Phys. Rev. D 73, 057301 (2006)
\hspace*{\fill}  hep-ph/0601116 (\version)
\vspace*{2\baselineskip}
\title{Possible new source of T and CP violation in neutrino oscillations}

\author{Frans R.\ Klinkhamer}
\email{frans.klinkhamer@physik.uni-karlsruhe.de}
\affiliation{Institute for Theoretical Physics,
University of Karlsruhe (TH), 76128 Karlsruhe, Germany
\vspace*{2\baselineskip}
}

\begin{abstract}
\vspace*{.5\baselineskip}\noindent  
A model is presented to illustrate that vacuum neutrino oscillations
can be essentially \DS{T} and \DS{CP} invariant up to a certain energy
but strongly \DS{T} and \DS{CP} noninvariant at much higher energies.
Detailed model results for the vacuum probabilities $P(\nu_\mu \to \nu_e)$
and \mbox{$P(\nu_e \to \nu_\mu)$} are given, which may be relevant to
proposed long-baseline neutrino-oscillation \mbox{experiments}.
\vspace*{2\baselineskip}
\end{abstract}

\maketitle

\section{Introduction}
\label{sec:Introduction}

One of the main goals of neutrino-oscillation physics in the
coming decennia will be to determine (or constrain) the violation
of time-reversal (\DS{T}) invariance and charge-conjugation--parity
(\DS{CP}) invariance in the lepton sector; see, e.g.,
Refs.~\cite{Apollonio-etal2002,KayserSSI2004}
for a comprehensive report and recent review article.
With three-flavor neutrino oscillations
being solely due to mass differences \cite{BilenkyPontecorvo1978},
the ultimate source of this \DS{T} and \DS{CP} violation would be the
complex Dirac phase $\delta$ in the unitary mixing matrix (here, denoted $X$)
between weak-interaction states and mass states \cite{MNS1962+KM1973}.

Another possible contribution to neutrino oscillations may come
from Lorentz-noninvariant Fermi-point-splitting effects
\cite{KlinkhamerJETPL2004,KlinkhamerIJMPA2006,KlinkhamerPRD2005}.
(The Fermi-point-splitting mechanism of neutrino oscillations
has a direct motivation from condensed-matter physics
\cite{KlinkhamerVolovikJETPL2004,KlinkhamerVolovikIJMPA,KlinkhamerVolovikJETPL2005},
but there have been many other suggestions for alternative mechanisms; see, e.g.,
Ref.~\cite{Hooper-etal} for an extensive list of references.)
With Fermi-point splittings present,
there is then a new unitary mixing matrix ($Y$) between
weak-interaction states and Fermi-point states.
If there are both mass differences and Fermi-point-splittings in
the neutrino sector, the relevant mixing matrix for neutrino oscillations
is between weak-interaction states and \emph{propagation} states,
where the neutrino propagation is affected simultaneously by mass and
Fermi point. This mixing matrix ($Z$) is determined, in
part, by the matrices $X$ and $Y$ of the mass and Fermi-point sectors,
respectively.

The crucial point, now, is that the Fermi-point-splitting matrix $Y$ may
have mixing angles ($\chi_{ij}$) and complex Dirac phase ($\omega$)
completely different from those of the mass-sector matrix $X$
(usually, denoted $\theta_{ij}$ and $\delta$).
In particular, there is the possibility that \emph{all} parameters
$\chi_{ij}$ and $\omega$ are nonvanishing, or even maximal.
This would then correspond to a new source
of \DS{T} (and \DS{CP}) violation effects in neutrino oscillations.
The goal of the present article is to illustrate this possibility
with a relatively simple model.

A potential new source of leptonic \DS{T} and \DS{CP} violation is all
the more interesting as neutrinos may play a decisive role in the creation
of the observed matter--antimatter asymmetry of the universe
\cite{FukugitaYanagida1986,Branco-etal2002,Buchmuller-etal2005}.
The suggestion is that
neutrinos would in some way be responsible for the creation of a
net lepton number $L$ at very high temperature
($T \gg M_W \approx 10^{2}\;\mathrm{GeV}$), which, at the
electroweak scale ($T \sim M_W$), is partially transformed
by sphaleron processes into a net baryon number $B$
\cite{tHooftPRL1976,Christ1980,KlinkhamerManton1984,KRS1985,KlinkhamerLee2001}.
Even though it will be difficult to relate the ultrahigh-energy
\DS{CP} violation needed for leptogenesis
to any \DS{T} and \DS{CP} violation of neutrino-oscillation
experiments at relatively low energies \cite{Branco-etal2002}
and  the fundamental mechanism of electroweak
$B+L$ violation at high temperatures ($T \gtrsim M_W$)
is not fully understood \cite{KlinkhamerLee2001},
the topic of leptonic \DS{T} and \DS{CP} violation
can be expected to play an important role in a discussion of
the physics of the early universe.

The outline for the remainder of this article is as follows.
In Sec.~\ref{sec:Model}, we describe the model.
In Sec.~\ref{sec:Results}, we give model results for vacuum
oscillation probabilities in the so-called ``golden channel,''
$\nu_e\leftrightarrow \nu_\mu$.
In Sec.~\ref{sec:Conclusion}, we present concluding remarks.

\section{Model}
\label{sec:Model}

\subsection{General remarks}
\label{sec:Model-GeneralRemarks}

In a previous article \cite{KlinkhamerPRD2005}, we have considered
a simple three-flavor neutrino-oscillation model
with both mass-square differences ($\Delta m^2_{ij}$)
and timelike Fermi-point splittings ($\Delta b_0^{(ij)}$).
The mixing of the mass sector was taken to be bi-maximal and
the one of the Fermi-point-splitting sector trimaximal, with
all complex phases vanishing.
The model had furthermore a hierarchy of
Fermi-point splittings ($b_0^{(1)} = b_0^{(2)} \ne b_0^{(3)}$)
which parallels the hierarchy of mass squares ($m^2_1 = m^2_2 \ne m^2_3$).
For the physics motivation of this type
of model (e.g., quantum phase transitions in superfluids), see
Refs.~\cite{KlinkhamerVolovikJETPL2004,KlinkhamerVolovikIJMPA,KlinkhamerVolovikJETPL2005}
and references therein.
As to the expected energy scale of neutrino Fermi points, there are
speculations \cite{KlinkhamerVolovikIJMPA,endnote-hierarchies}
but no firm predictions.

The present article extends the previous one by
presenting results on the appearance probability
$P_{\mu e}\equiv P(\nu_\mu\rightarrow \nu_e)$ from a
generalized model with the same mass hierarchy as the model of
Ref.~\cite{KlinkhamerPRD2005} but with equidistant Fermi-point splittings
($\,b_0^{(2)}- b_0^{(1)}=b_0^{(3)}- b_0^{(2)}\,$)
and one nonvanishing complex phase ($\omega =\pi/4$).
In addition, we will consider the case of relatively strong Fermi-point-splitting
effects compared to mass-difference effects,
whereas Ref.~\cite{KlinkhamerPRD2005}
focused on relatively weak splitting effects. For this purpose,
we introduce a new parametrization (with nonnegative dimensionless parameters
$\rho$ and $\tau$) which makes a straightforward comparison
between different long-baseline neutrino-oscillation experiments possible.
Relatively weak or strong Fermi-point-splitting effects then correspond
to $\tau \ll 1$ or $\tau \gtrsim 1$, respectively.
The behavior of $P_{\mu e}(\rho,\tau)$ turns out to be quite
complicated for $\tau \gtrsim 1$.

For this generalized model with complex phase $\omega =\pi/4$,
we also give the model probability
of the time-reversed process, $\nu_e \rightarrow \nu_\mu$.
It will be seen that the generalized model has a rather interesting
phenomenology with stealthlike characteristics in certain cases
and strong time-reversal noninvariance in others.

In this article, we mainly speak about possible \DS{T}--violating effects
in neutrino oscillations from Fermi-point splitting.
Whether or not there are corresponding \DS{CP}--violating effects
depends on the (unknown) physics responsible for
the Fermi-point splittings, i.e., whether or not there is \DS{CPT} invariance.
Depending on the Fermi-point splittings of the
right-handed ``antineutrinos'' compared to those of the left-handed ``neutrinos,''
there may or may not be \DS{CP} violation in addition to the  \DS{T} violation
of the model considered ($\sin\omega\ne 0$); see
Sec.~4 of Ref.~\cite{KlinkhamerIJMPA2006} for further details.
For the rest of this article, we take an agnostic point of view on the
\DS{CPT} invariance of Fermi-point splitting, and focus on the manifest
\DS{T} violation from the presence of complex phases in the Hamiltonian.

\subsection{Specifics}
\label{sec:Model-Specifics}

Setting  $\hbar=c=1$ and writing $p \equiv |\mathbf{p}|$ for the (large)
neutrino momentum, the Hamiltonian of the generalized version of the model
of Ref.~\cite{KlinkhamerPRD2005} contains
three terms in the $(\nu_e,\nu_\mu,\nu_\tau)$ flavor basis,
\beq
H \supset p\,\openone
+  X\cdot  D_{m} \cdot  X^{\dagger}
+ D_{\alpha\beta}\cdot Y\cdot  D_{b_0}\cdot Y^{\dagger}\cdot D^{\dagger}_{\alpha\beta}
\;,
\label{3flavormatrix}
\eeq
with diagonal matrices
\begin{subequations}
\beqa
\hspace*{-5mm}
  D_{m}   &\equiv&   \mathrm{diag} \big(\,
m^2_{1}/(2 p) \, , \,
m^2_{2}/(2 p) \, , \,
m^2_{3}/(2 p)
\,\big) \, , \\[2mm]
\hspace*{-5mm}
 D_{b_0} &\equiv&  \mathrm{diag} \big(\,
b_0^{(1)}  \, , \,
b_0^{(2)}  \, , \,
b_0^{(3)}
\,\big) \, , \\[2mm]
\hspace*{-5mm}
D_{\alpha\beta} &\equiv& \mathrm{diag} \big(
\exp[\i\alpha]\,, \, \exp[-\i(\alpha+\beta )]\, , \, \exp[\i\beta]
\,\big)\,,
\eeqa
\end{subequations}
and $SU(3)$ matrices
\begin{subequations}
\beqa
X &\equiv& M_{32}(\theta_{32}) \cdot M_{13}(\theta_{13},\delta)
           \cdot M_{21}(\theta_{21})
\,,\\[2mm]
Y &\equiv& M_{32}(\chi_{32}) \cdot M_{13}(\chi_{13},\omega)
         \cdot M_{21}(\chi_{21})\,,
\label{XYdef}
\eeqa
\end{subequations}
in terms of the basic  matrices
\begin{subequations}
\beqa
\hspace*{-5mm}
M_{32}(\vartheta) &\equiv&
\threemat{1 &\;\;\; 0 \;\;\;& 0 \\
          0 &\;\;\; \cos\vartheta \;\;\;& \sin\vartheta \\
          0 &\;\;\; -\sin\vartheta \;\;\;&\cos\vartheta}
\,,\\[2mm]
\hspace*{-5mm}
M_{21}(\vartheta) &\equiv&
\threemat{\cos\vartheta &\;\;\; \sin\vartheta \;\;\;& 0 \\
          -\sin\vartheta&\;\;\; \cos\vartheta \;\;\;& 0 \\
         0 & 0 & 1}
\,,\\[2mm]
\hspace*{-5mm}
M_{13}(\vartheta,\varphi) &\equiv&
\threemat{\cos\vartheta & \;\;\;0\;\;\; & \;e^{i\varphi}\,\sin\vartheta \\
          0 & \;\;\;1\;\;\; & 0 \\
          -e^{-i\varphi}\,\sin\vartheta& \;\;\;0\;\;\; & \cos\vartheta}
\,.
\label{Mdef}
\eeqa
\end{subequations}
The following dimensionless parameters are chosen in the mass sector:
\begin{subequations}
\label{fixedparameters}
\beqa
\hspace*{-9mm}
R_m &\equiv& \Delta m^2_{21}/\Delta m^2_{32}
   \equiv  ( m_2^2-m_1^2)/ (m_3^2-m_2^2) =0\,,
\label{fixedparameters-r-Mass} \\[2mm]
\hspace*{-9mm}
\theta_{21}&=&\theta_{32}=\pi/4 \,,\quad\theta_{13}=0\,,
\quad \delta = 0\,,
\label{fixedparameters-anglesphaseMass}
\eeqa
in the Fermi-point-splitting sector:
\beqa
\hspace*{-8mm}
R&\equiv& R_{b_0} \equiv \frac{\Delta b_0^{(21)}}{\Delta b_0^{(32)}}
  \equiv \frac{b_0^{(2)}- b_0^{(1)}}{b_0^{(3)}- b_0^{(2)}} \in (-\infty,\infty)\,,
\label{fixedparameters-R-FPS} \\[2mm]
\hspace*{-8mm}
\chi_{21} &=&\chi_{32}=\chi_{13} =\pi/4\,,\quad \omega\in [0,2\pi) \,,
\label{fixedparameters-anglesphaseFPS}
\eeqa
and for the relative complex phases between mass and Fermi-point sectors:
\beqa
\alpha &=& \beta=  0\,.
\label{fixedparameters-phases}
\eeqa
\end{subequations}
In addition to the dimensionless parameters $R$ and $\omega$,
there are two dimensionful model parameters relevant to neutrino oscillations,
\beq
\Delta m^2_{31}   \equiv m_3^2-m_1^2           >0 \,,\quad
\Delta b_0^{(31)} \equiv b_0^{(3)}- b_0^{(1)}  >0\,,
\label{freeparameters}
\eeq
which have been taken positive. Remark that the mass-sector parameters
(\ref{fixedparameters}ab) are not unrealistic
(with $\sin\theta_{13}=0$, the chosen value of $\delta$ is, in fact, irrelevant)
but the sign of
$\Delta m^2_{31}$ is still undetermined experimentally
\cite{Apollonio-etal2002,KayserSSI2004}.

For high-energy neutrino oscillations over a travel distance $L$,
two dimensionless parameters can be defined as follows ($E_\nu \sim p$):
\begin{subequations} \label{rhotau}
\beqa
\rho &\equiv&
  \fracnew{2\,E_\nu\,\hbar c}{L\,|\Delta m^2_{31}|\,c^4} \approx
  1.5786 \; \Bigg(\, \fracnew{E_\nu}{10\;\mathrm{GeV}} \,\Bigg)
 \Bigg(\, \fracnew{10^{3}\;\mathrm{km}}{L} \,\Bigg)\,
 \Bigg(\, \fracnew{2.5\times 10^{-3}\;\mathrm{eV}^2/c^4}{|\Delta m^2_{31}|}\,\Bigg)\,,
\label{rho}\\[3mm]
\tau  &\equiv&
  \fracnew{L\,|\Delta b_0^{(31)}|}{\hbar c} \approx
  5.0671 \; \Bigg(\, \fracnew{L}{10^{3}\;\mathrm{km}} \,\Bigg)
  \Bigg(\, \fracnew{|\Delta b_0^{(31)}|}{10^{-12}\;\mathrm{eV}} \,\Bigg)   \,,
\label{tau}
\eeqa
\end{subequations}
with $\hbar$ and $c$ temporarily reinstated.
These two dimensionless parameters, together with  $R$ and $\omega$,
completely determine the oscillation probabilities, at least for the simple
model considered and with matter effects  neglected \cite{endnote-matter}.

\section{Results}
\label{sec:Results}

\begin{figure*}[p]   
\vspace*{-.5cm}
\begin{center}\includegraphics[width=14cm]{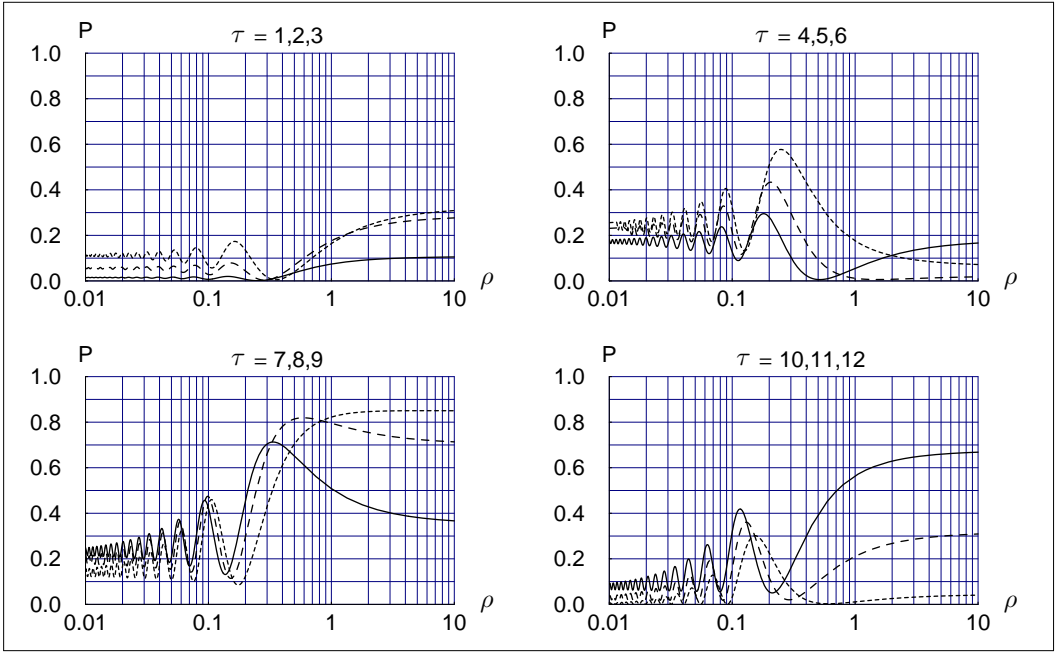}
\end{center}
\vspace*{-.75cm}
\caption{Numerical model results for the vacuum neutrino-oscillation
probability $P_{\mu e} \equiv P(\nu_\mu\rightarrow \nu_e)$
as a function of the dimensionless parameters $\rho$ and $\tau$,
defined by Eqs.~(\ref{rhotau}ab).
The model, described in Sec.~\ref{sec:Model-Specifics}, has
Fermi-point-splitting ratio $R =1$ and complex phase $\omega =\pi/4$.
Shown are constant--$\tau$ slices of $P(\rho,\tau) \equiv P_{\mu e}(\rho,\tau)$,
where the solid, long-dashed, and short-dashed curves correspond to
$\tau = 1,2,0 \pmod 3$, respectively.}
\label{figPmuTOe-R1-omegaPi/4}
\end{figure*}

The model defined by Eqs.~\eqref{3flavormatrix}--\eqref{freeparameters},
for $R=1$ and $\omega=\pi/4$,
gives the vacuum probability $P_{\mu e} \equiv P(\nu_\mu\rightarrow \nu_e)$
shown in Fig.~\ref{figPmuTOe-R1-omegaPi/4} as a function of the
parameters $\rho$ and $\tau$ from Eqs.~(\ref{rhotau}ab).
For $\rho \to \infty$ (high neutrino energies at fixed $\Delta m^2_{31}\times L$),
this model is similar to the pure Fermi-point-splitting
model studied previously \cite{KlinkhamerJETPL2004,KlinkhamerIJMPA2006},
for which $P_{\mu e}$ is known exactly \cite{endnote-FPSmodels}.
For $\rho \to 0$ (low energies), the $P_{\mu e}$ behavior can also be understood
analytically \cite{endnote-deg-pert-theory}.

These last remarks explain the observed stealthlike behavior
of $P_{\mu e}(\rho,\tau)$ at certain special values of $\tau$, with
the appearance probability being nonzero only for a relatively small range of energies.
An example would be given by the case of  $\tau \approx 12$ in
Fig.~\ref{figPmuTOe-R1-omegaPi/4}.
For a given nonzero value of $\Delta b_0^{(31)}$, the particular
appearance probability  $P_{\mu e}$ would be nearly shut off
at the corresponding distance $L \approx 12/|\Delta b_0^{(31)}|$,
reappearing, however, at generic values of $L$ \cite{endnote-stealth}.

\begin{figure*}[p]   
\begin{center}\includegraphics[width=14cm]{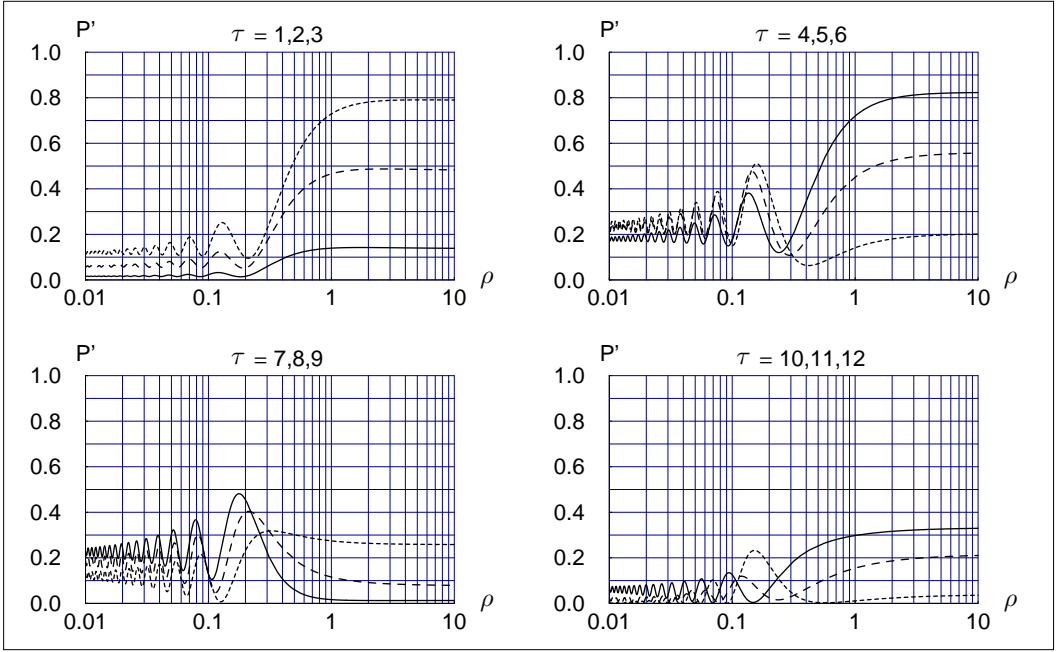}
\end{center}
\vspace*{-.75cm}
\caption{Same as Fig.~\ref{figPmuTOe-R1-omegaPi/4}
but for the time-reversed process with probability
$P^\prime \equiv P(\nu_e\rightarrow \nu_\mu)$. If \DS{CPT} invariance holds,
$P^\prime$ also corresponds to $P(\overline{\nu}_\mu\rightarrow \overline{\nu}_e)$.}
\label{figPeTOmu-R1-omegaPi/4}
\vspace*{-0.25cm}
\end{figure*}

The model with $R=1$ and $\omega=\pi/4$
can be expected to have \DS{T} violation for large enough
neutrino energy. (The corresponding pure mass-difference model,
relevant at low energies,
does not have \DS{T} violation, in particular,
because $\sin\theta_{13}$ vanishes.)
The probability of the $\nu_e \rightarrow \nu_\mu$ process
is shown in Fig.~\ref{figPeTOmu-R1-omegaPi/4} and
the difference with Fig.~\ref{figPmuTOe-R1-omegaPi/4},
for $\rho \gtrsim 0.2$ and generic values
of $\tau$, indeed signals time-reversal noninvariance.

If \DS{CPT} invariance holds true (cf. Sec.~\ref{sec:Model-GeneralRemarks}),
the curves of Fig.~\ref{figPeTOmu-R1-omegaPi/4}
also apply to $P(\overline{\nu}_\mu\rightarrow \overline{\nu}_e)$
and the difference with Fig.~\ref{figPmuTOe-R1-omegaPi/4} signals
\DS{CP} noninvariance.
If, on the other hand, \DS{CPT} invariance is violated maximally,
the probability $P(\overline{\nu}_\mu\rightarrow \overline{\nu}_e)$
is given by the curves of Fig.~\ref{figPmuTOe-R1-omegaPi/4}
and there is only \DS{T} violation, with $P(\nu_\mu\rightarrow \nu_e)$ $=$
$P(\overline{\nu}_\mu\rightarrow \overline{\nu}_e)$ $\ne$
$P(\nu_e \rightarrow \nu_\mu)$.

%

%

\section{Conclusion}
\label{sec:Conclusion}

Figures~\ref{figPmuTOe-R1-omegaPi/4} and \ref{figPeTOmu-R1-omegaPi/4}
show strong time-reversal (\DS{T}) noninvariance of
vacuum neutrino oscillations $\nu_\mu \leftrightarrow \nu_e$
at high energies, which traces back to the large complex
phase $\omega$ of the model considered,
together with the large mixing angles $\chi_{ij}$ and ratio $R$ in the
Fermi-point-splitting sector \cite{endnote-SK}.
(As mentioned in the last paragraph of Sec.~\ref{sec:Model-GeneralRemarks},
there may or may not be a corresponding \DS{CP} violation, depending
on whether or not the Fermi-point splittings respect \DS{CPT}.)
In other words, this \DS{T}  (and \DS{CP}) violation would primarily
take place \emph{outside} the mass sector and show up at the high
end of the neutrino energy spectrum \cite{endnote-intermediateTviolation}.

A neutrino factory
with broad energy spectrum $E_\nu \approx 10-50\;\mathrm{GeV}$ and
several detectors at baselines $L$ up to $12800\;\mathrm{km}$
would be the ideal machine, in principle, to establish such
strong \DS{T} (and \DS{CP}) violation in high-energy neutrino
oscillations \cite{Geer1997,Cervera-etal2000,Bueno-etal2001,Apollonio-etal2002}.
Perhaps nearer in the future,  (redesigned) superbeam experiments such as
NO$\nu$A \cite{NOvA2005} can also look for
possible new sources of \DS{CP} violation. High-energy cosmic
neutrinos might provide additional information; cf. Ref.~\cite{Hooper-etal}.

As mentioned in the Introduction, \emph{any} new source
of \DS{T} (and \DS{CP}) violation in neutrino oscillations,
especially at the high end of the neutrino energy spectrum,
may be relevant to
the physics of the early universe. Moreover, this new
\DS{T} (and \DS{CP}) violation
could be related to the emergence of the \SM~from a nonrelativistic
underlying theory. The search for new sources of \DS{T} and \DS{CP} violation
in high-energy neutrino oscillations is therefore an important task of future
superbeam and neutrino-factory experiments.

\noindent\section*{ACKNOWLEDGMENTS}

It is a pleasure to thank Gustavo Branco, Jacob
Schneps, and Grisha Volovik for useful discussions.


\begin{thebibliography}{99}
\bibitem{endnote-intermediateTviolation}
Model \eqref{3flavormatrix} can also have \DS{T} and  \DS{CP} violation
at intermediate neutrino energies, for example, by
setting $\delta=\omega=0$ and having generic $\alpha$ and $\beta$.
\end{thebibliography}
\end{document}